\documentclass[10pt, prd, amssymb,aps,twocolumn]{revtex4}
\usepackage{graphics}
\usepackage{bm}

\begin{document}
\draft

\title{How Phase Transitions induce classical
behaviour}

\author{R.J. Rivers$^{1}$}
\email{r.rivers@imperial.ac.uk}
\author{F.C. Lombardo$^{2}$}%
\email{lombardo@df.uba.ar}

\affiliation{$^{1}$Theoretical Physics Group; Blackett Laboratory,
Imperial College, London SW7 2BZ} \affiliation{$^2$ Departamento
de F\'\i sica, Facultad de Ciencias Exactas y Naturales,
Universidad de Buenos Aires - Ciudad Universitaria, Pabell\' on I,
1428 Buenos Aires, Argentina}
\date{\today}

\begin{abstract}

We continue the analysis of the onset of classical behaviour in a
scalar field after a continuous phase transition, in which the
system-field, the long wavelength order parameter of the model,
interacts with an environment, of its own short-wavelength modes
and other fields, neutral and charged, with which it is expected
to interact. We compute the decoherence time for the system-field
modes from the master equation and directly from the decoherence
functional (with identical results). In simple circumstances the
order parameter field is classical by the time the transition is
complete.

\end{abstract}
\maketitle
03.70.+k,  05.70.Fh, 03.65.Yz

\section{Introduction}
The standard big bang cosmological model of the early universe
assumes a period of rapid cooling, giving a strong likelihood of
phase transitions, at the grand unified and electroweak scales
\cite{Kolb} in particular.

In this talk we describe how phase transitions naturally take us
from a quantum to classical description of the universe.
Metaphysics aside, cosmologists rely on the fact that the relevant
fields obey classical equations from early times, since it is not
possible to solve the quantum theory directly. Fortunately, we
have reason to believe that (continuous) transitions will move us
rapidly to classical behaviour. Classical behaviour arises in the
following way:
 \begin{itemize}
 \item
Classical correlations: By this is meant that the Wigner
function(al) $W[\pi ,\phi]$ peaks on classical phase-space
trajectories, with a probabilistic interpretation.
 \item
 Diagonalisation: By this is meant that the density matrix $\rho
 (t)$ should become (approximately) diagonal, in this case in a
 field basis. Alternatively, we can demand diagonalisation of the
 decoherence functional. In either case a probabilistic description (no
 quantum interference)
  is obtained.
  \item
  Stochastic behaviour: The decoherence functional, which provides
  the diffusion (noise) to diagonalise the density matrix also supplies
  the dissipation that enables the fields to obey probabilistic
  stochastic equations,
  which evolve into classical equations.
 \end{itemize}

\noindent From the papers of Guth and Pi \cite{guthpi} onwards, it
has been appreciated that unstable modes lead to classical
correlations through squeezing. On the other hand, we understand
diagonalisation to be an almost inevitable consequence of tracing
over the 'environment' of the 'system' modes.

Continuous transitions supply both ingredients, from which the
classical equations follow. Firstly, the field ordering after such
a transition is due to the growth in amplitude of unstable
long-wavelength modes, which arise automatically from unstable
maxima in the potential. Secondly, the stable short-wavelength
modes of the field, together with all the other fields with which
it interacts, form an environment whose coarse-graining enforces
diagonalisation and makes the long-wavelength modes decohere.

What matters are the time scales. An ideal situation, which we
shall show is possible, is that the theory becomes classical in
the sense above, before the transition is complete. However, to
quantify this is difficult because, with fields, we are dealing
with infinite degree of freedom systems. One of us (F.L) has shown
elsewhere \cite{diana} how classical correlations arise in quantum
mechanical systems that mimic the field theory that we shall
consider here, and we refer the reader to that paper for the role
that classical correlations play.  Our concern in this talk is,
rather, with diagonalisation, determined both through the master
equation for the evolution of the density matrix and the
decoherence functional, whose role is to describe consistent
histories. Stochastic equations are then a corollary to this same
diagonalisation.

This talk builds upon earlier published work by us and Diego
Mazzitelli \cite{lomplb,lomplb2,lomnpb}, together with our
contributions to the proceedings of the 2001 meeting in Peyresq
\cite{P1,P2} and we refer the reader to this earlier work for much
of the basic technical details. We restrict ourselves to flat
space-time. The extension to non-trivial metrics is
straightforward in principle. See the recent work of Lombardo
\cite{POFer}, which complements this. The developments since the
last proceedings are our greater understanding of the use of trial
configurations and slower quenches \cite{lomnpb}, the parallel use
of the decoherence functional to characterise decoherence, and the
extension of the theory to include electromagnetism \cite{ubaIC}.

\section{Evolution of the density matrix}

 The evolution of a quantum field as it falls out of equilibrium at
a transition is determined in large part by its behaviour at early
times, before interactions have time to take effect. To be
concrete, consider a real scalar field $\phi (x)$, described by a
$Z_2$-symmetry breaking action ($\mu^2 >0$)
\begin{equation}
S[\phi ] = \int d^4x\left\{ \frac{1}{2}\partial_{\mu}
\phi\partial^{\mu} \phi + \frac{1}{2}\mu^2 \phi^2 -
\frac{\lambda}{4!}\phi^4\right\},
 \label{phi4}
\end{equation}
with symmetry breaking scale $\eta^2 = 6\mu^2/\lambda$. On
heating, this shows a continuous transition, with critical
temperature $T_{\rm c}^2 = 2\eta^2$. If, by virtue of the
expansion of the universe the system is very rapidly cooled
(quenched) from $T > T_{\rm c}$ to $T < T_{\rm c}$, the initial
stages of the transition can be described by a free field theory
with inverted mass $-\mu^2<0$. The state of the field is initially
concentrated on the local maximum of the potential, and spreads
out with time. This description is valid for short times, until
the field wave functional explores the ground states of the
potential.

The $\phi$-field ordering after the transition is due to the
growth in amplitude of its {\it unstable} long-wavelength modes,
which we term $\phi_<(x)$. For an instantaneous quench these have
wave-number $k<\mu$ for all time. Although the situation is more
complicated for slower quenches, until the transition is complete
there are always unstable modes. As a complement to these, we
anticipate that the {\it stable} short-wavelength modes of the
field $\phi_>(x)$, where
 \begin{equation}
 \phi (x) = \phi_<(x) + \phi_>(x),
 \nonumber
 \end{equation}
will form an environment whose coarse-graining makes the
long-wavelength modes decohere \cite{fernando}. In practice, the
boundary between stable and unstable is not crucially important,
provided there is time enough for the power in the field
fluctuations to be firmly in the long-wavelength modes. This
requires weak coupling $\lambda\ll 1$.  Of course, all the other
fields with which $\phi$ interacts will contribute to its
decoherence, but for the moment we ignore such fields (before
returning to them in the last section).

After splitting, the action (\ref{phi4}) can be written as
\begin{equation}S[\phi] = S[\phi_<] + S[\phi_>] + S_{\rm int}[\phi_<,
\phi_>],\label{actions}\end{equation} where the interaction term
is dominated \cite{lomplb,lomnpb} by its biquadratic term
\begin{equation}
S_{\rm int}[\phi_<, \phi_>] \approx -\frac{1}{6}\lambda \int d^4x
~ \phi_<^2(x) \phi_> ^2(x).
 \label{inter}
 \end{equation}

The total density matrix (for the system and bath fields) is
defined by
\begin{equation} \rho_{\rm r}[\phi^+,\phi^-,t]=
\rho[\phi_<^+,\phi_>^+,\phi_<^-,\phi_>^-,t]=\langle\phi_<^+
\phi_>^+\vert {\hat\rho} \vert \phi_<^-
\phi_>^-\rangle,\label{matrix} \nonumber
\end{equation}
and we assume that, initially, the thermal system and its
environment are not correlated.

On tracing out the short-wavelength modes, the reduced density
matrix
\begin{equation}\rho_{\rm r}[\phi_<^+,\phi_<^-,t] = \int {\cal D}\phi_>
\rho[\phi_<^+,\phi_>,\phi_<^-,\phi_>,t],\label{red}\nonumber
\end{equation}
whose diagonalisation determines the onset of classical behaviour,
evolves as
\begin{equation}\rho_{\rm r}[t] = \int d\phi_{<\rm i}^+
\int d\phi_{<\rm i}^- ~ J_{\rm r}[t,t_{\rm i}]~ \rho_{\rm
r}[t_{\rm i}],\label{evol}\nonumber
\end{equation}
where $J_{\rm r}[t,t_{\rm i}]$ is the evolution operator
\begin{equation}J_{\rm r}[t,t_{\rm i}] =
\int_{\phi_{<\rm i}^+}^{\phi_{<\rm f}^+}{\cal D}\phi_<^+
\int_{\phi¯_{<\rm i}^-}^{\phi_{<\rm f}^-}{\cal D}\phi_<^-
\exp\{iS_{CG}[\phi_<^+,\phi_<^-]\}.\label{evolred}
\end{equation}
$S_{CG}[\phi_<^+,\phi_<^-]$ is the coarse-grained effective
action, of the closed time-path form
\begin{equation}S_{CG}[\phi_<^+,\phi_<^-] = S[\phi_<^+] - S[\phi_<^-] + \delta
S[\phi_<^+,\phi_<^-].\label{CTPEA}\nonumber\end{equation}
 All the
information about the effect of the environment is encoded in
$\delta S[\phi_<^+,\phi_<^-]$ through the influence functional (or
Feynman-Vernon functional \cite{feynver})
\begin{equation}F[\phi_<^+,\phi_<^-] = \exp \{i
\delta S[\phi_<^+,\phi_<^-]\}.\label{IA}\nonumber\end{equation}

$\delta S$ has a well defined diagrammatic expansion, of the form
\begin{eqnarray}
&&\delta S[\phi_<^+,\phi_<^-]= \langle S_{\rm
int}[\phi_<^+,\phi_>^+]\rangle - \langle
S_{\rm int}[\phi_<^-,\phi_>^-]\rangle \nonumber \\
&& +{i\over{2}}\{ \langle S_{\rm int}^2[\phi_<^+,\phi_>^+]\rangle
-
\big[\langle S_{\rm int}[\phi_<^+,\phi_>^+]\rangle \big]^2 \} \nonumber \\
&&- i \langle S_{\rm int}[\phi_<^+,\phi_>^+] S_{\rm
int}[\phi_<^-,\phi_>^-]\rangle  \nonumber \\
&&+ i \langle S_{\rm int}[\phi_<^+,\phi_>^+]\rangle \langle
S_{\rm int}[\phi_<^-,\phi_>^-]\rangle \nonumber  \\
&&+{i\over{2}}\{S^2_{\rm int}[\phi_<^-,\phi_>^-]\rangle -
\big[\langle S_{\rm
int}[\phi_<^-,\phi_>^-]\rangle\big]^2\}.\label{inflac}
\end{eqnarray}
The quantum averages of the functionals of the fields are with
respect to the free field action of the environment, defined as
\begin{eqnarray}
&&\langle B[\phi_>^+,\phi_>^-] \rangle = \int d\phi_{>\rm i}^+
\int d\phi_{>\rm i}^-
~ \rho_{>}[\phi_{>\rm i}^+,\phi_{>\rm i}^-,t_0] \nonumber \\
&& \times \int d\phi_{>\rm f}^+ \int_{\phi_{>\rm i}^+}^{\phi_{>\rm
f}^+}{\cal
D}\phi_>^+\int_{\phi_{>\rm i}^-}^{\phi_{>\rm f}^-}{\cal D}\phi_>^- \nonumber \\
&& \times \exp{{i\over{\hbar}}\{S_0[\phi_>^+] - S_0[\phi_>^-]\}}
B[\phi_>^+,\phi_>^-].\label{averag}\nonumber\end{eqnarray}

To lowest order $\lambda^2$ diagrams are one-loop in the short
wavelength modes.

\subsection{The Master Equation}

Once the reduced density matrix has become approximately diagonal
quantum interference has effectively disappeared and the density
matrix permits a conventional probability interpretation. To see
how the diagonalisation of $\rho_{\rm r}$ occurs, we construct the
{\it master equation}, which casts its evolution in differential
form.
 As a first approximation, we make a
saddle-point approximation for $J_{\rm r}$ in Eq.(\ref{evolred}),
\begin{equation}
J_{\rm r} [\phi_{<{\rm f}}^+,\phi_{<{\rm f}}^-, t_f\vert
\phi_{<{\rm i}}^+,\phi_{<{\rm i}}^-, t_i] \approx\exp (i
S_{CG}[\phi^+_{<\rm cl},\phi^-_{<\rm cl}]), \label{saddle}
\end{equation}
In (\ref{saddle}) $\phi^\pm_{<\rm cl}$  is the solution to the
equation of motion
\begin{equation}
\frac{\delta Re S_{CG}}{\delta\phi_<^+}\bigg |_{\phi_<^+=\phi_<^-}
=0\nonumber,
\end{equation}
 with boundary conditions $\phi^\pm_{\rm cl}(t_0)=\phi^\pm_{<\rm i}$
and $\phi^\pm_{<\rm cl}(t)=\phi^\pm_{<\rm f}$.

It is very difficult to solve this equation analytically. We
exploit the fact that, even if the universe is completely
homogeneous prior to the transition then, after the transition,
causality requires \cite{kibble} that it be inhomogeneous because
of the finite speed at which the order parameter fields can order
themselves.  This is in contra-distinction to the usual adiabatic
analysis in which (for the continuous transition that interest us
here) the correlation length diverges at the transition.

Since the field cannot be homogeneous in either of its
groundstates $\phi = \eta$ or $\phi = -\eta$ there is an effective
'domain' structure in which the domain boundaries are 'walls'
across which $\phi$ flips from one groundstate to the other.
Further, these domains have a characteristic size $\xi$, where
$\xi^{-1}= \pi k_0$ labels the dominant momentum in the power of
the $\phi$-field fluctuations as the unstable long-wavelength
modes grow exponentially. For simplicity, we adopt a
'minisuperspace' approximation, in which we assume regular
domains, enabling $\phi_{<\rm cl}(\vec x, s)$ to be written as
\begin{equation}
\phi_{<\rm cl}(\vec x, s) =  f(s,t)\Phi (x)\Phi (y)\Phi (z),
\label{phiclass}\end{equation}
 where
$\Phi (0) = \Phi (\xi) = 0$, and
\[\Phi (x+\xi) = -\Phi (x).
 \label{phicl}
\]
\begin{figure}
\scalebox{0.50}{\includegraphics{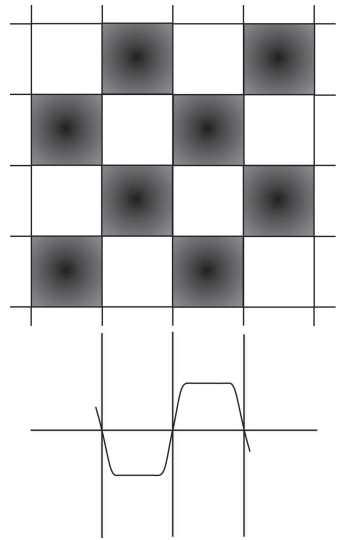}} \caption{The field
profile (\ref{phiclass}) in two dimensions, with lattice size
$\xi$. Dark areas represent $\phi\approx\eta$, light areas
$\phi\approx -\eta$. The boundaries are domain walls, with profile
given below.} \label{additive}
\end{figure}
$f(s,t)$ satisfies $f(0,t)= \phi_{<\rm i}$ and $f(t,t) =
\phi_{<\rm f}$. We write it as
\begin{equation} f(s,t) = \phi_{<\rm i} u_1(s,t) +
\phi_{<\rm f} u_2(s,t).
 \label{us}
 \end{equation}
 In \cite{lomnpb} we made the simplest choice for $\Phi(x)$,
 $$ \Phi(x) = \cos k_0x.$$
 Extensions to include more Fourier modes
 are straightforward in principle, but our work in \cite{lomnpb} was sufficient
 to show that the results only depend weakly on the details of the domain function
 $\Phi(x)$ for few Fourier modes.
 In the light of the more
 qualitative comments made here, we refer the reader again to
 \cite{lomnpb} for details.
On the other hand, the $u_i(s,t)$ are solutions of the mode
equation for wavenumber $k_0$ during the quench, with boundary
conditions $u_1(0,t) = 1$, $u_1(t,t) = 0$ and $u_2(0,t) = 0$,
$u_2(t,t) = 1$.

In order to obtain the master equation we must compute the final
time derivative of the propagator $J_{\rm r}$. After that, all the
dependence on the initial field configurations $\phi^\pm_{\rm i}$
(coming from the classical solutions $\phi^\pm_{\rm cl}$) must be
eliminated. Assuming that the unstable growth has implemented
diagonalisation before back-reaction is important, $J_{\rm r}$ can
be determined, approximately, from the {\it free} propagators as
\begin{equation}J_0[t,t_{\rm i}]=
\int_{\phi_{<\rm i}^+}^{\phi_{<\rm f}^+}{\cal D}\phi_<^+
\int_{\phi¯_{<\rm i}^-}^{\phi_{<\rm f}^-}{\cal D}\phi_<^-\exp\{i[
S_0(\phi_<^+) - S_0(\phi_<^-)]\}\label{propdeJ0}
\end{equation}
where $S_0$ is the free-field action. This  satisfies the general
identities \cite{fernando}
\begin{equation}
\phi^\pm_{\rm cl}(s) J_0 = \Big[\phi^\pm_{\rm f} [u_2(s) -
\frac{{\dot u}_2(t)}{{\dot
             u}_1(t)}u_1(s)] \mp  i {u_1(s)\over{{\dot u}_1(t)}}
\partial_{\phi^\pm_{<\rm f}}\Big]J_0
\label{rel1}\nonumber
\end{equation}
which allow us to remove the initial field configurations
$\phi^\pm_{\rm i}$, and obtain the master equation.

Even with these simplifications the full equation is very
complicated, but it is sufficient to calculate the correction to
the usual unitary evolution coming from the noise (diffusion)
kernels (to be defined later). The result reads
\begin{equation}
i {\dot \rho}_{\rm r} = \langle \phi^+_{<\rm f}\vert [H,\rho_{\rm
r}] \vert \phi^-_{<\rm f}\rangle - i V \Delta^2 D(\omega_0, t)
\rho_{\rm r}+ ... \label{master}
\end{equation}
where $D$ is the diffusion coefficient and $$\Delta = (\phi_{<\rm
f}^{+2} - \phi_{<\rm f}^{-2})/2$$ for the {\it final} field
configurations (henceforth we drop the suffix). The ellipsis
denotes other terms coming from the time derivative that do not
contribute to the diffusive effects. $V$ is understood as the
minimal volume inside which there are no coherent superpositions
of macroscopically distinguishable states for the field.

\subsection{The diagonalisation of $\rho_{\rm r}$}

 The
effect of the diffusion coefficient in driving the diagonaliation
can be seen by considering the following approximate solution to
the master equation:
\begin{equation} \rho_{\rm r}[\phi_<^+, \phi_<^-; t]\approx
\rho^{\rm u}_{\rm r}[\phi_<^+, \phi_<^-; t] ~ \exp \left[-V
\Delta^2 \int_0^t ds ~D(k_0, s) \right],\nonumber
\end{equation} where $\rho^{\rm u}_{\rm r}$ is the solution of the
unitary part of the master equation (i.e. without environment).
The system will decohere when the non-diagonal elements of the
reduced density matrix are much smaller than the diagonal ones.

The decoherence time  $t_{D}$ sets the scale after which we have a
classical system-field configuration, and depends strongly on the
properties of the environment. It satisfies
\begin{equation}
1 \approx  V\Delta^2\int_{0}^{t_{\rm D}} ds ~D(k_0,s) ,
\label{Dsum}
\end{equation}
and corresponds to the time after which we are able to distinguish
between two different field amplitudes, inside a given volume $V$.

 To terms up to order $\lambda^2$ and one loop in
the $\hbar$ expansion (we continue to work in units in which
$\hbar = k_B = 1$), the influence action due to the biquadratic
interaction between system and environment has real imaginary
parts
\begin{equation}  {\rm Re}\delta S=
\int d^4x\int d^4y ~\Delta (x)K(x,y)\Sigma (y), \label{inff}
\end{equation}
and
\begin{equation}
{\rm Im}\delta S  = - \frac{1}{2}\int d^4x \int d^4y
 \Delta(x)N(x,y) \Delta
(y), \label{ImDS}
\end{equation}
where $K(x,y)=\frac{1}{2}\lambda^2 \theta (t - t')
 {\rm Im}G^{> 2}_{++}(x-y)$ is the dissipation kernel and $N(x,y) =
\frac{1}{2}\lambda^2 {\rm Re}G^{>2}_{++}(x,y)$ is the noise
(diffusion) kernel. $G^{>}_{++}(x, y)$ is the thermal
short-wavelength closed time-path correlator. The UV singular
parts of the loop diagrams are implicitly removed by
renormalisation, leaving the finite temperature parts which are
$O(T^2)$.
 We also have defined
\begin{equation}
\Sigma =\frac{1}{2}(\phi_<^{+2} + \phi_<^{-2})
\end{equation}
for final state modes.

 Explicit calculation shows that $D(k_0,t)$
is built from the diffusion kernel $N$ as
\begin{equation} D(k_0,t)
= \int_0^t ~ ds ~ u(s,t)~ F(k_0,s,t) \label{D0}
\end{equation}
where
$$u(s,t) =\bigg[u_2(s,t) - \frac{{\dot u}_2(t,t)}{{\dot
u}_1(t,t)}u_1(s,t)\bigg]^{2}.$$

$ F(k_0,s,t)$ is constructed from the spatial Fourier transforms
of the overlap of the diffusion kernel with the field profiles
$\Phi (x)\Phi (y)\Phi (z)$. For the single mode $ \Phi(x) = \sin
k_0x$
\begin{eqnarray}
&&F(k_0,s,t)= \frac{\lambda^2}{64}[{\rm Re}G^{>2}_{++}(0; t-s) \nonumber \\
&&+ \frac{3}{2}{\rm Re} G^{>2}_{++}(2k_0; t-s)
+ \frac{3}{4}{\rm Re} G^{>2}_{++}(2\sqrt{2}k_0; t-s) \nonumber \\
&&+\frac{1}{8}{\rm Re} G^{>2}_{++}(2\sqrt{3}k_0; t-s)].
\end{eqnarray}
In the integrand of (\ref{D0}) $u(s,t)$ is rapidly varying, driven
by the unstable modes, and $F(k_0,s,t)$ is slowly varying. For
long-wavelengths $k_0\ll\mu$ we have, approximately,
$$ F(k_0,s,t) = O(N(k_0 =0; t-s)),$$ whereby
\begin{equation}
 D(k_0,t) \approx   F(k_0, 0,t) ~
\int_0^{t} ds ~ u(s,t). \label{D02}
\end{equation}
That is, the diffusion coefficient factorises into the
environmental term $F$, relatively insensitive to both wavenumber
and time, and the rapidly growing integral that measures the
classical growth of the unstable system modes that are ordered in
the transition.

To be specific, we restrict ourselves to the simplest case of an
instantaneous quench from a temperature $T = {\cal O}(T_{\rm c}) >
T_{\rm c}$, for which
 \begin{equation}
 u_1=  {\sinh[\omega_0 (t -
s)] \over{\sinh(\omega_0 t)}},\,\,u_2(s,t)=  {\sinh(\omega_0 s)
\over {\sinh(\omega_0 t),}},\,\,\, \label{us2}
\end{equation}
where $\omega_0^2 = \mu^2 - k_0^2\approx \mu^2$. It follows that
\begin{equation}
u(s,t) = \cosh^2[\omega_0(t-s)],
\end{equation}
from whose end-point behaviour at $s=0$ of the integral
(\ref{D02}) we find the even simpler result
\begin{equation}
D(k_0,t)\sim \mu^{-1}F(k_0, 0,t)~ u(0,t) \sim (\lambda T_{\rm
c}/4\pi\mu)^2~ \exp [2\mu t], \label{D(t)}
\end{equation}
assuming $\mu t_D\gg 1$. The ${\cal O}(T_{\rm c}^2)$ behaviour of
$F$ derives from the thermal short-wavelength modes.

For more general quenches growth is more complicated than simple
exponential behaviour but a similar separation into fast and slow
components applies.

We have omitted a large amount of complicated technical detail
(see \cite{lomnpb}), to give such a simple final result. This
suggests that we could have reached the same conclusion more
directly.

We now indicate how we can obtain the same results by demanding
consistent histories of the $\phi$ field.

\section{The Decoherence Functional}

The notion of consistent histories provides a parallel approach to
classicality. Quantum evolution can be considered as a coherent
superposition of fine-grained histories. If one defines the
c-number field $\phi (x)$ as specifying a fine-grained history,
the quantum amplitude for that history is $\Psi [\phi] \sim
e^{iS[\phi]}$ (we continue to work in units in which $\hbar =1$).

In the quantum open system approach that we have adopted here, we
are concerned with coarse-grained histories
\begin{equation}
\Psi [\alpha] = \int {\cal D}\phi ~ e^{iS[\phi]}\alpha [\phi]
\end{equation}
where $\alpha [\phi]$ is the filter function that defines the
coarse-graining.

From this we define the decoherence function for two
coarse-grained histories as
\begin{equation}
 {\cal D}[\alpha^+,\alpha^-] = \int {\cal D}\phi^+{\cal
 D}\phi^-~e^{i(S[\phi^+]-S[\phi^-])}\alpha^+ [\phi^+]\alpha^-
 [\phi^-].
\end{equation}
${\cal D}[\alpha^+,\alpha^-]$ does not factorise because the
histories $\phi^{\pm}$ are not independent; they must assume
identical values on a spacelike surface in the far future.
Decoherence means physically that the different coarse-graining
histories making up the full quantum evolution acquire individual
reality, and may therefore be assigned definite probabilities in
the classical sense.

A necessary and sufficient condition for the validity of the sum
rules of probability theory (i.e. no quantum interference terms)
is \cite{Gri}
\begin{equation}
 {\rm Re}{\cal D}~[\alpha^+,\alpha^-]\approx 0,
\end{equation}
when $\alpha^+\neq\alpha^-$ (although in most cases the stronger
condition ${\cal D}[\alpha^+,\alpha^-]\approx 0$ holds
\cite{Omn}). Such histories are consistent \cite{GH}.

For our particular application, we wish to consider as a single
coarse-grained history all those fine-grained ones where the full
field $\phi$ remains close to a prescribed classical field
configuration $\phi_{\rm cl}$. The filter function takes the form
\begin{equation}
 \alpha_{\rm cl}[\phi ] = \int {\cal D}J~
 e^{i\int J(\phi - \phi_{\rm cl})}\alpha_{\rm cl}[J].
\end{equation}
In the general case, $\alpha[\phi]$ is a smooth function (we
exclude the case $\alpha[\phi]=$ const, where there is no
coarse-graining at all). Using

\begin{equation}J\phi \equiv \int d^4x J(x) \phi (x),
\end{equation}
we may write the decoherence functional between two classical
histories as

\begin{eqnarray}
{\cal D}[\alpha^+,\alpha^-] &=& \int {\cal D}J^+{\cal
 D}J^-~e^{i W[J^+,J^-] - (J^+ \phi_{\rm cl}^+ - J^- \phi_{\rm cl}^-)}\nonumber \\
&\times & \alpha^+[J^+]\alpha^{-*}[J^-],
\end{eqnarray}
where

\begin{equation}
e^{i W[J^+,J^-]} = \int  {\cal D}\phi^+ {\cal D}\phi^- ~
e^{i(S[\phi^+] - S[\phi^-] + J^+\phi^+ - J^-\phi^-)}
,\end{equation} is the closed-path-time generating functional.

In principle, we can examine general classical solutions for their
consistency but, in practice, it is simplest to restrict ourselves
to solutions of the form (\ref{phiclass}). In that case, we have
made a de facto separation into long and short-wavelength modes
whereby, in a saddle-point approximation over $J$. In this way, we
can see that the above expression is formally equivalent to the
definition of the influence functional (see Ref. \cite{fernando}
for details). Thus, we may write
\begin{equation}
 {\cal D}(\phi^+_{\rm cl},\phi^-_{\rm cl}) \sim
 \exp \{iS_{CG}[\phi^+_{\rm cl},\phi^-_{\rm cl}]\}.
\end{equation}
As a result,
 \begin{equation}
|{\cal D}(\phi^+_{\rm cl},\phi^-_{\rm cl})| \sim
 \exp \{-{\rm Im}\delta S[\phi^+_{\rm cl},\phi^-_{\rm cl}]\}
 \end{equation}
For the instantaneous quench of (\ref{us2}), using the late time
behaviour $\phi_{\rm cl}^{\pm}\sim e^{\mu s}\phi_0^{\pm}$, ${\rm
Im}\delta S[\phi^+_{\rm cl},\phi^-_{\rm cl}]$ takes the form
\begin{equation}
{\rm Im}~\delta S \sim \frac{V\Delta^2}{\mu^2}\int_0^t ds\int_0^t
ds' e^{2\mu s}~e^{2\mu s'} F(k_0,s,s'). \label{ImS}
\end{equation}

From this viewpoint adjacent histories become consistent at the
time $t_D$, for which
\begin{equation}
 1\approx \int_0^{t_D} dt ~{\rm Im~\delta S}.
\label{tD2}
\end{equation}

At this level, after performing the stationary phase
approximation, it is equivalent to evaluate the decoherence time
scale from the master equation (through diffusion terms) or
directly from the decoherence functional (or the influence
functional).

\section{The decoherence time}

We have used the same terminology for the time $t_D$ since, on
inspection, (\ref{tD2}) is {\it identical} to (\ref{Dsum}) in
defining the onset of classical behaviour. As we noted, in
practice the use of the decoherence functional looks to be  less
restrictive than the master equation, and we hope to show this
elsewhere.

For the moment what is of interest is whether $t_D$, based on
linearisation of the model, occurs before backreaction sets in, to
invalidate this assumption. When all the details are taken into
account, whether from (\ref{us}) or (\ref{us2}), $t_D$ satisfies
\begin{equation}
1= {\cal O}\bigg(\frac{\lambda^2VT_{\rm
c}^2}{\mu^3}\Delta^2\bigg)~\exp(4\mu t_D),
 \label{tD3}
\end{equation}
or, equivalently
\begin{equation}
\exp(4\mu t_D) = {\cal O}\bigg(\frac{\mu^3}{\lambda^2VT_{\rm
c}^2\Delta^2}\bigg) .\end{equation} For the rapid quenches
considered here, linearisation manifestly breaks down by the time
$t^*$, for which $\langle \phi^2\rangle_{t^*} \sim \eta^2$, given
by
\begin{equation}
\exp (2\mu t^*) =  {\cal O}\bigg( \frac{\mu}{\lambda T_{\rm
c}}\bigg).
 \label{tstar}
\end{equation}
The exponential factor, as always, arises from the growth of the
unstable long-wavelength modes. The factor $T_{\rm c}^{-1}$ comes
from the $\coth(\beta\omega /2)$ factor that encodes the initial
Boltzmann distribution at temperature $T\gtrsim T_{\rm c}$.

Our conservative choice is that the volume factor $V$ is ${\cal
O}(\mu^{-3})$  since $\mu^{-1}$ (the Compton wavelength) is the
smallest scale at which we need to look. With this choice it
follows that
\begin{equation}
\exp 2(t^* - t_D) = {\cal O}\bigg(\frac{|\Delta|}{\mu^2}\bigg)) =
{\cal O}(\bar\phi\delta), \label{tstartD}
\end{equation}
where $\bar\phi = (\phi_{<}^+ + \phi_{<}^-)/2\mu,$ and $ \delta =
|\phi_{<}^+ - \phi_{<}^-|/2\mu$. Within the volume $V$ we do not
discriminate between field amplitudes which differ by $ {\cal
O}(\mu) $, and therefore take $\delta = {\cal O}(1)$. For
$\bar\phi$ we note that, if $t_D$ were to equal $t^*$, then
$\bar\phi^2 = {\cal O}(1/\lambda)={\cal O}(T_{\rm c}^2/\mu^2)\gg
1,$ and in general $\bar\phi
>1$. As a result, if there are no large numerical factors, we have
\begin{equation}
t_D < t^*, \label{tllt}
\end{equation}
and the density matrix has become diagonal before the transition
is complete.  Detailed calculation shows \cite{lomnpb} that there
are no large factors \cite{footnote}.

We already see a significant difference between the behaviour for
the case of a biquadratic interaction  with an environment given
by (\ref{Dsum}) and the more familiar linear interaction, adopted
because it can be solvable (e.g. \cite{kim}). This latter would
have replaced $\Delta/\mu^2$ just by $\delta$, incapable of
inducing decoherence before the transition is complete. Although
linear environments can be justified in quantum mechanics, in
quantum field theory a purely linear environment corresponds to an
inappropriate digonalisation of the action.

We note that, once the interaction strength is sufficiently weak
for classical behaviour to appear before the transition is
complete, this persists, however weak the coupling becomes. It
remains the case that, the  weaker the coupling, the longer it
takes for the environment to decohere the system but, at the same
time, the longer it takes for the transition to be completed, and
the ordering (\ref{tllt}) remains the same. This is equally true
for more general quenches provided the system remains
approximately Gaussian until the transition is complete.

\subsection{Back-reaction}

In both calculations for the decoherence time we have been obliged
to assume that free-field behaviour explains the exponential
growth of the long-wavelength modes. In reality, we are thinking
of $\phi_{<\rm f}$ as describing the symmetry-broken phase, with
magnitude $\eta$, the symmetry breaking scale (if we normalise
$|\Phi (x)|$ to be unity at its maxima). It can be shown
\cite{Karra} that, for an instantaneous quench at least, nonlinear
behaviour that stops the exponential growth only becomes important
just before $t^*$. To see this, we adopt the Hartree
approximation, in which the equations of motion are linearised
self-consistently. With a little work we find that the theory only
ceases to behave like a free Gaussian theory with upside-down
potential at a time $t_B$, where
\begin{equation}
t^* - t_B = O(\mu^{-1}). \label{tB}
\end{equation}
It follows that $t_B\geq t_D$ in our ordering of scales $T_{\rm
c}\gg\mu$.

\section{Late-time behaviour}

When (\ref{tllt}) is valid, we see that $\rho_{\rm r}$ becomes
diagonal before non-linear terms could be relevant. Although we
haven't discussed it here,  classical behaviour has been achieved
before quantum effects can destroy the positivity of the Wigner
function $W_{\rm r}$, which is enforced by the unstable modes.
Really, our $t_{\rm D}$ sets the time after which we have a
classical probability distribution (positive definite) even for
times $t > t_B$. The existence of the environment is crucial in
doing this.

This result also justifies in part the use of phenomenological
stochastic equations to describe the dynamical evolution of the
system field, as we will now discuss. As it is well known
\cite{fernando,greiner}, one can regard the imaginary part of
$\delta S$ as coming from a noise source $\xi (x)$, with a
Gaussian functional probability distribution.
\begin{equation}
P[\xi (x)]= {\cal N}_{\xi} \exp\bigg\{-{1\over{2}}\int d^4x\int
d^4y ~\xi (x) N^{-1}\xi (y)\bigg\},
\end{equation}
where ${\cal N}_{\xi}$ is a normalization factor.  This enables us
to write the imaginary part of the influence action as a
functional integral over the Gaussian
field $\xi (x)$ %
\begin{eqnarray}
&&\int {\cal D}\xi (x) P[\xi ]\exp{\left[ -i \bigg\{\int
d^4x~\Delta (x) \xi (x) \bigg\}\right]}\nonumber \\ &=&
\exp{\bigg\{-\frac{i}{2}\int d^4x\int d^4y \ \Big[\Delta (x)
~N(x,y)~ \Delta (y)\Big] \bigg\}}.\end{eqnarray} %
In consequence, the coarse-grained effective action can be rewritten as %
\begin{equation}S_{CG}[\phi_<^+,\phi_<^{-}]=-{1\over{i}} \ln  \int {\cal D} \xi
  P[\xi ]
\exp\bigg\{i S_{\rm eff}[\phi_<^{+},\phi_<^{-}, \xi ]\bigg\},
\end{equation}
where
\begin{equation}
S_{\rm eff}[\phi_<^{+},\phi_<^{-},\xi ]= {\rm Re}
S_{CG}[\phi_<^{+},\phi_<^{-}]- \int d^4x\Big[\Delta (x) \xi (x)
\Big].
\end{equation}

The functional variation equation%
\begin{equation}
\left.{\delta S_{\rm eff}[\phi_<^+,\phi_<^-, \xi_{2}]\over{\delta
\phi_<^+}}\right\vert_{\phi_<^+=\phi_<^-}=0, \label{statphase}
\end{equation}
``semiclassical-Langevin'' equation for the system-field
\cite{fernando,greiner}
\begin{eqnarray}
\left.{\delta {\rm Re}S_{CG}[\phi_<^+,\phi_<^-,
\xi_{2}]\over{\delta \phi_<^+}}\right\vert_{\phi_<^{\pm}
=\phi_<}=\xi (x)\phi_<. \label{lange2}
 \end{eqnarray}

The evolution equation for the reduced Wigner functional $W_r$ now
becomes the Fokker-Planck counterpart to (\ref{lange2}).

Each part of the environment that we include leads to a further
'dissipative' term on the left hand side of (\ref{lange2}) with a
countervailing noise term on the right hand side.  Although the
$\phi_<\phi^3_>$ and $\phi^3_<\phi_>$ terms were ignorable in the
bounding of $t_D$, in the Langevin equations they give further
terms, with quadratic $\phi_<^2\xi_3$ noise and linear (additive)
noise $\xi_1$ respectively.

For times later than $t_B$, neither perturbation theory nor more
general non-Gaussian methods are valid. It is difficult to imagine
an {\it ab initio} derivation of the dissipative and noise terms
from the full quantum field theory. In this sense, a reasonable
alternative is to analyze phenomenological stochastic equations
numerically and check the robustness of the predictions against
different choices of the dissipative kernels and of the type of
noise. Hitherto, pure {\it additive} noise has been the basis for
empirical stochastic equations in relativistic field theory that
confirm Kibble's causal analysis \cite{laguna}. However, recent
numerical simulations with a more realistic mix of additive and
multiplicative noise has shown that domain formation is unchanged
\cite{nunopedroray}.

\section{Further environments: Neutral Fields}

Finally, it has to be said that taking only the short wavelength
modes of the field as a one-loop system environment is not a
robust approximation. This is particularly so for the Langevin
equation (\ref{lange2}) \cite{gleiser}. We should be summing over
hard thermal loops in the $\phi$-propagators. To be in proper
control of the diffusion we need an environment that interacts
with the system, without the system having a strong impact on the
environment. This requires us to introduce further deconfining
environments. We are helped in that, in the early universe, the
order parameter field $\phi$ will interact with any field $\chi$
for which there is no selection rule. Again, it is the biquadratic
interactions that are the most important.

The most simple additional environment is one of a large number
$N\gg 1$ of weakly coupled scalar fields $\chi_{\rm a}$, for which
the action (\ref{phi4}) is extended to

\begin{equation}
S[\phi , \chi ] = S[\phi ] + S[\chi ] + S_{\rm int}[\phi ,\chi ],
\label{quaction}
\end{equation}
where $S[\phi]$ is as before, and
\begin{eqnarray}
&&S[\chi_{\rm a} ] = \sum_{\rm a=1}^N\int d^4x\left\{
{1\over{2}}\partial_{\mu}\chi_{\rm a}
\partial^{\mu}
\chi_{\rm a} - {1\over{2}} m_{\rm a}^2 \chi^2_{\rm a}\right\},
\nonumber
\\
&&S_{\rm int}[\phi ,\chi ] = - \sum_{\rm a=1}^N\frac{g_{\rm a}}{8}
\int d^4x \phi^2 (x) \chi^2_{\rm a} (x), \label{Sint}
\end{eqnarray}
where $m_{\rm a}^2 >0$. For simplicity we take weak couplings
$\lambda \simeq g_{\rm a}$ and comparable masses $m_{\rm a}\simeq
\mu$. The effect of a large number of weakly interacting
environmental fields is twofold. Firstly, the $\chi_a$ fields
reduce the critical temperature $T_c$ and, in order that $T_{\rm
c}^2=\frac {2\mu^2}{\lambda + \sum g_{\rm a}}\gg \mu^2$, we must
take $\lambda + \sum g_{\rm a} \ll 1$.  Secondly, the single
$\chi$-loop contribution to the diffusion coefficient is the
dominant $\chi$-field effect if, for order of magnitude estimates,
we take identical $g_{\rm a} =\bar g/\sqrt{N}$, whereby $1\gg
1/\sqrt N\gg\bar g\simeq \lambda$. With this choice the effect of
the $\phi$-field on the $\chi_a$ thermal masses is, relatively,
$O(1/\sqrt{N})$ and can be ignored. We stress that this is not a
Hartree or large-N approximation of the type that, to date, has
been the main way to proceed\cite{boya,mottola,Greg} for a {\it
closed} system.

Provided the change in temperature is not too slow the exponential
instabilities of the $\phi$-field grow so fast that the field has
populated the degenerate vacua well before the temperature has
dropped to zero. Since the temperature $T_c$ has no particular
significance for the environment field, for these early times we
can keep the temperature of the environment fixed at $T_{\chi}
={\cal O}(T_{\rm c})$ (our calculations are only at the level of
orders of magnitude). As before, we split the field as $\phi =
\phi_< + \phi_>$. The $\chi$-fields give an additional one-loop
contribution to $D(k_0, t)$ with the same $u(s)$ but a $G_{++}$
constructed from (all the modes of) the $\chi$-field. The
separation of the diffusion coefficient due to $\chi$ into fast
and slow factors proceeds as before to give a term that is
identical to (\ref{D(t)}) (or (\ref{ImS})) but for its ${\bar
g}^2$ prefactor.

Diffusion effects are {\it additive} at the one-loop level, and
the final effect is to replace $\lambda^2$ in (\ref{tD3}) by
$\lambda^2 + {\bar g}^2 >\lambda^2$, while leaving (\ref{tstar})
unchanged. Although the relationship between $T_c$ and $\lambda$
has been uncoupled by the presence of the $\chi_a$, the
relationship (\ref{tstartD}) persists, with an enhanced right hand
side, requiring that (\ref{tllt}) is even better satisfied.

\section{Charged fields}

Given that the effect of further environmental fields is to
increase the diffusion coefficient and speed up the onset of
classical behaviour, additional fields interacting with the $\phi$
field seem superfluous. However, the symmetries of the universe
seem to be local (gauge symmetries), rather than global, and we
should take gauge fields into account. We conclude with some
observations from our work in progress \cite{ubaIC} with local
symmetry breaking.

Local symmetry breaking is not possible for our real $\phi$ field
but, as a first step \cite{lomplb2}, it is not difficult  to
extend our model to that of a complex $\phi$-field. At the level
of $O(2)$ global interactions with external fields and with its
own short-wavelength modes, everything goes through essentially as
before. The main difference is in the choice of single degree of
freedom configurations.  Writing
$$\phi(x) = \frac{1}{\sqrt{2}}(\phi_1(x) +
i\phi_2(x)),$$
 we assume that the $\phi_a$ behave independently until back-reaction is important.
The simplest single-mode approximation to the long-wavelength
system field is
\begin{equation}
\phi_{1,<\rm cl}(\vec x, s) =  f_1(s,t)\Phi (x)\Phi (y)\Phi (z),
\label{phi1}
 \end{equation}
say, and
\begin{equation}
\phi_{2,<\rm cl}(\vec x, s) =  f_2(s,t)\Phi (x+a)\Phi (y+b)\Phi
(z+c), \label{phi2}
 \end{equation}
for some non-zero $a,b,c$.
 $f_a(s,t)$ satisfies $f_a(0,t)= \phi_{a,\rm i}$ and $f_a(t,t) = \phi_{a,\rm
f}$. We write them as
\begin{equation} f_a(s,t) = \phi_{a,\rm i} u_1(s,t) +
\phi_{a,\rm f} u_2(s,t),
 \label{us3}
 \end{equation}
 as before. Whereas the classical mode (\ref{phiclass}) of the
 real scalar described a regular array of domain walls, separation
 $\xi$, defined by the zeroes $\phi_{<\rm cl}(\vec x) = 0$,
 the  complex $\phi_{<\rm cl}(\vec x)$
 describes a regular array of {\it line} zeroes (the intersections of
 $\phi_{1<\rm cl}(\vec x) = 0 =\phi_{2<\rm cl}(\vec
 x)$), which will evolve into global vortices after the
 transition. Although our assumption of a regular lattice of
 vortices is an extreme simplification, the production of vortices
 with typical separation $\xi$ is as we would expect
 \cite{kibble}.

 In fact, to date we have not even been as sophisticated as
 (\ref{phi1}) and (\ref{phi2}), but have just taken periodicity in
 a single direction \cite{P2}. This is sufficient to see that the
 system decoheres before the transition is complete, with an almost identical relation (\ref{tstartD}). We assume
 that the insensitivity of the prefactor $F(k_0,s,t)$ to the regular lattice in both
 (\ref{D0}) and
 (\ref{ImS}) is equally true here. This will be examined
 elsewhere \cite{ubaIC}.

 Local $U(1)$ symmetry breaking is most
easily accommodated by taking the $\phi$-field to interact with
other charged fields $\chi$ through the local $U(1)$ action
\begin{equation}
S [\phi, A_{\mu},\chi ] = S [\phi, A_{\mu}] + S_{\chi}
[A_{\mu},\chi ],
 \label{S}
\end{equation}
in which $S[\phi, A_{\mu} ] =$
\begin{equation}
\int d^4x\left\{(D_{\mu} \phi)^*D^{\mu} \phi + \mu^2 \phi^*\phi -
{\lambda\over{4}}(\phi^*\phi)^2
-\frac{1}{4}F^{\mu\nu}F_{\mu\nu}\right\}, \label{SQED}
\end{equation}
and
\begin{equation}
S [A_{\mu}, \chi ] = \int d^4x\left\{(D_{\mu} \chi)^*D^{\mu} \chi
+ m^2\chi^*\chi\right\}. \label{chi}
\end{equation}
We have taken a single $\chi$ field. The theory (\ref{S}) shows a
phase transition, and we assume couplings are such as to make this
transition continuous.

For simplicity, let us just take $\chi$ to be the environment to
the system field $\phi$, which we do not separate into short and
long-wavelength modes. On integrating out the $\chi$-field
environment, the reduced density matrix $\rho_{{\rm r}}[\phi^+,
A_{\mu}^+,\phi^-, A_{\mu}^-, t]$ evolves as
\begin{equation}\rho_{\rm r}[t] = \int d\phi_{\rm i}^+
\int d\phi_{\rm i}^-\int dA_{\rm i}^+\int dA_{\rm i}^- ~ J_{\rm
r}[t,t_{\rm i}]~ \rho_{\rm r}[t_{\rm
i}].\label{evolgauge}\nonumber
\end{equation}
(We have dropped the indices on $A_{\mu}$ for clarity). Yet again,
we make a saddle-point approximation,
\begin{equation}
J_{\rm r} [t_f, t_{\rm i}] \approx\exp (i S_{CG}[\phi^+_{\rm cl},
A^+_{\rm cl},\phi^-_{\rm cl}. A^-_{\rm cl}]), \label{saddlegauge}
\end{equation}
where the coarse-grained action $S_{CG}$ has the form
\begin{eqnarray}
S_{CG}[\phi^+,A^+,\phi^-, A^-]=&& S[\phi^+, A^+] - S[\phi^-,A^-]\nonumber\\
&&  + \delta S[\phi^+,A^+,\phi^-, A^-]. \label{CTPE}
\end{eqnarray}
As before, $\delta S$ encodes all the interactions between the
environment and the system.
 In (\ref{saddlegauge}) $\phi^\pm_{\rm cl}$  is the solution
to the equation of motion
\begin{equation}
\frac{\delta {\rm Re} S_{CG}}{\delta\phi^+} = \frac{\delta {\rm
Re} S_{CG}}{\delta A^+}  =0 \end{equation}
 subject to
$\phi^+=\phi^-$ and $A^+ = A^-$, with boundary conditions
$\phi^\pm_{\rm cl}(t_0)=\phi^\pm_{\rm i}$ and $\phi^\pm_{\rm
cl}(t)=\phi^\pm_{\rm f}$, and similarly for $A^{\pm}_{\rm cl}$.

We stress that we are not tracing over the electromagnetic degrees
of freedom, but determining the indirect effect of the $\chi$
environment on the $\phi$ field, mediated by electromagnetism.

Again, for simplicity, we assume an instantaneous quench. The
diffusion is again driven by the unstable $\phi$ modes that,
approximately as
\begin{equation}
(\Box - \mu^2)\phi_{\rm cl}(s,{\bf x}) = (\Box - \mu^2)\phi_{\rm
cl}^*(s,{\bf x}) = 0 \label{phic2}
\end{equation}
for times $s\lesssim t^*$. This unstable scalar $\phi_{\rm cl}$ is
the source for the classical electromagnetic field, $A_{\rm
cl}^{\mu}(s,{\bf x})$, satisfying
\begin{eqnarray}
&&\partial^{\nu}F_{\nu\mu}(s,{\bf x})+ (e^2|\phi_{\rm cl}|^2 + e^2
G_{++}(0))A_{\mu, \rm cl}(s,{\bf x})\nonumber\\
&& + e^2\int_0^s \,dt\,\int\,d^3y\, {\rm Im} \Delta_{\mu\nu}(s-t,
{\bf
x}-{\bf y} )A_{\rm cl}^{\nu}(s,{\bf y})\nonumber\\
&& = j_{\mu}(s,{\bf x}), \label{Acl}
\end{eqnarray}
in the Lorentz gauge, where $j_{\mu} = -ie\phi^*_{\rm
cl}\partial^\leftrightarrow_{\mu}\phi_{\rm cl}$.

In (\ref{Acl}), $G_{++}(x-y) = Tr [T(\chi
(x)\chi^{\dagger}(y)\rho_{\chi}(0)]$ is the hot $\chi$ propagator
at temperature $T$. The $G_{++}(0)$ term is the $\chi$-loop
thermal mass contribution to the $A_{\mu}$ field.

We interpret (\ref{Acl}) as being the start of an expansion with
solution
\begin{equation}
A_{\mu ,\rm cl}(s,{\bf x}) = \int\,d^3y\,dt\, D_{\mu\nu}(s,t;{\bf
x}-{\bf y})j^{\nu}(t,{\bf y}), \label{Acl2}
\end{equation}
where $D_{\mu\nu}(s,t;{\bf x}-{\bf y})$ is the thermal
$A_{\mu}$-field propagator in the $\chi$ heatbath. We have ignored
the oscillatory solution of $A_{\mu}$ to the homogeneous equation,
since this will not induce the exponentially growing diffusion
that we need for rapid decoherence.

Just as for the other models considered earlier, when the
transition is completed, there is a characteristic scale, the
separation of the local vortices that express the frustration die
to causal bounds. If we adopt a single characteristic scale before
then, ${\rm Im}\delta S$ now has two contributions. We have
already seen that the first, of the form (\ref{ImDS}), but from
the $\chi$-loop, is sufficient to enforce decoherence before the
transition is complete, for acceptable parameters. We also have a
contribution of the form \cite{ubaIC}
\begin{equation}
{\rm Im}\delta S  = \frac{-e^2}{2}\int d^4x \int d^4y
 (\Delta A)^{\mu}_{(x)}{\rm Re}\Delta_{\mu\nu}(x,y) (\Delta A)^{\nu}
_{(y)}, \label{ImDS2}
\end{equation}
due to the electromagnetic field, where $(\Delta A)^{\mu}= A_{\rm
cl}^{+\mu}- A_{\rm cl}^{-\mu}$, derived from $\phi$ through
(\ref{Acl2}), and
 \vspace{0.2cm}
\begin{eqnarray}
\Delta_{\mu\nu}(x-y) &=& \bigg(\frac{\partial}{\partial z^{\mu}
}-\frac{\partial}{\partial
x^{\mu}}\bigg)\bigg(\frac{\partial}{\partial w^{\mu}
}-\frac{\partial}{\partial
y^{\mu}}\bigg)\nonumber \\
&\times & G_{++}(x-w)G_{++}(z-y)\bigg|^{z=x}_{w=y} \label{DA}
\end{eqnarray}

This additional term to the diffusion function has derivative
couplings. Having made a gauge choice, these give rise to explicit
momenta factors $k_{\mu}$ in the generalisation of $F$. Unlike the
contributions to $D$ that we have seen so far, which are largely
insensitive to the momentum scale $k_0$, these contributions are
strongly damped at large wavelength. In consequence, it is likely
that they barely enhance the onset of classical behaviour but,
given that the effect of the other environmental modes is to
enforce classical behaviour so quickly, it hardly matters. We
intend to give a fuller discussion of this elsewhere \cite{ubaIC}.

\section{Conclusion}

We have shown how, for fast quenches, weakly coupled environments
make a scalar order parameter field decohere before the transition
is complete, under very general assumptions. An essential
ingredient for rapid decoherence is {\it nonlinear} coupling to
the environment, inevitable when that environment contains the
short wavelength modes of the order parameter field. Had we only
considered linear coupling to the environment, as in \cite{kim},
for example (but an assumption that is ubiquitous in quantum
mechanical models, from Brownian motion onwards) decoherence would
not have happened before the transition was complete, and we would
not know how to proceed, although classical correlations would
have occurred.  For weak couplings further scalar environments
with local interactions with the system field only make
decoherence more rapid. However, it seems that, for the relevant
case of a charged environment, also interacting indirectly through
electromagnetic interactions, this indirect contribution has
little effect on a decoherence that is already effective.

\section{acknowledgments}
 We thank Diego Mazzitelli for his
collaboration in this work. F.C.L. was supported by Universidad de
Buenos Aires, CONICET (Argentina), Fundaci\'on Antorchas and
ANPCyT. R.J.R. was supported in part by the COSLAB programme of
the European Science Foundation.


\begin{references}
\bibitem{footnote} In evaluating $F$ we perform the
loop diagram using the full propagator rather than just its short
wavelength modes. Since the $T_{\rm c}$ behaviour comes entirely
from the short wavelength part of the integral, this is justified
\end{references}
\end{document}